# Deep multilevel wet etching of fused silica glass microstructures in BOE solution

*Konstantinova T.G.[1,2], Andronic M. M.[1], Baklykov D.A.[1], Stukalova V.E.[1], Ezenkova D.A.[1], Zikiy E.V.[1], Bashinova M.V.[1], Solovev A.A[1], Lotkov E.S.[1], Ryzhikov I.A.[1] and Rodionov I.A.[1,2, *]*

[1] *FMN Laboratory, Bauman Moscow State Technical University, 105005 Moscow, Russia, st. Rubtsovskaia nab. 2/18;*
[2] *Dukhov Automatics Research Institute, (VNIIA), 127055 Moscow, Russia, Suschevskaya Ul., 22*
**\*irodionov@bmstu.ru**

**Abstract**

Fused silica glass is a material of choice for micromechanical, microfluidic, and optical devices due to its ultimate chemical resistance, optical, electrical, and mechanical performance. Wet etching in hydrofluoric solutions especially a buffered oxide etching (BOE) solution is still the key method for fabricating fused silica glass-based microdevices. It is well known that protective mask integrity during deep fused silica wet etching is a big challenge due to chemical stability of fused glass and extremely aggressive BOE properties. Here, we propose a multilevel fused silica glass microstructures fabrication route based on deep wet etching through a stepped mask with just a one grayscale photolithography step. First, we provide a deep comprehensive analysis of a fused quartz dissolution mechanism in BOE solution and calculate the main fluoride fractions like $HF_2^-$, $F^-$, $(HF)_2$ components in a BOE solution as a function of pH and NH4F:HF ratio at room temperature. Then, we experimentally investigate the influence of BOE concentration (NH4F:HF from 1:1 to 14:1) on the mask resistance, etch rate and profile isotropy during fused silica 60 minutes etching through a metal/photoresist mask. Finally, we demonstrate a high-quality multilevel over-200 μm isotropic wet etching process with the rate up to 3 μm/min, which could be of a great interest for advanced fused silica microdevices with flexure suspensions, inertial masses, microchannels, and through-wafer holes.

**1. Introduction**

Fused silica glass wafers are widely used in micro-devices such as inertial sensors [1], microfluidic systems [2], and optical sensors [3] due to their excellent mechanical, electrical, and optical properties, thermal and chemical stability, as well as biocompatibility. The structural elements of these devices usually contain flexure membranes 5–50 μm thick, microchannels 10–100 μm deep, or through holes for the entire depth of the substrate from 150 μm to 1000 μm. In addition, micro-devices often combine these elements into multilevel microstructures. It is critically important to ensure high quality processing of fused glass microdevice elements, since it determine the optical, rheological, and mechanical parameters of the structures. There are three major glass microfabrication technologies: mechanical, thermal, and chemical (dry and wet) [4]. Only chemical methods ensure obtaining smooth surfaces, which are critical for various optical, mechanical and microfluidic applications. In contrast to wet methods, plasma etching is limited by low etch rate and depth of etching due to poor selectivity to protective masks [5]. That is why wet etching processes are still the key method of glass microdevices fabrication. It allows etching deep microstructures with an isotropic profile and low surface roughness at high etching rates (several μm/min) [6]. Fused silica etching is carried out in hydrofluoric acid (HF) solutions due to high chemical inertness of glass. Usually, buffer additives are added to hydrofluoric acid solutions to stabilize an etching rate, which is useful in the case of multicomponent glasses etching due to dissolution of reaction products [7]. However, protective mask stability and integrity in reactive etchants become the limiting factor (table 1).

Table 1 - Overview of protective mask materials

| Material | Disadvantages | Etch depth, μm | Sources |
|---|---|---|---|
| Photoresist | Low adhesion<br>High undercut | 20 - 80 | [2,8,9,10,11,12,13,14] |
| Cr | Low adhesion<br>High undercut<br>Pinholes formation | 30 - 100 | [15,16,17] |
| Cr/Au | High-cost<br>CMOS incompatibility<br>Pinholes formation | 5 - 500 | [12,18,19,20,21,22] |
| Cu | Very thick layers (>1um) | 100 – 250 | [12,23] |
| Mo | High stressed films | 200 – 250 | [24,25,26] |
| a:Si, bulk Si | High stressed films<br>Pinholes formation<br>High coating temperatures<br>KOH mask removal | 60 – 600 | [12,20,27,28,29] |

The material and properties of a protective mask, as well as composition of the etching solution, are the most important factors, which affect etching quality. Photoresists (AZ5214E, SPR220), metals (Au/Cr, Cr, Mo), and silicon-based (a:Si, bulk-Si) protective masks (Table 1) are the most commonly used solutions. The fabrication process complexity and required depths of fused silica microstructures determine the choice of mask materials for various devices. Thus, photoresist masks are easy to spin coat, but they have low adhesion and low resistance to HF solutions limiting etching depths at several tens of micrometers [8-14]. Si-based masks are highly resistant to hydrofluoric acid solution [20, 27-29]. However, the fabrication of low-stress Si-based layers is challenging (e.g., thick a:Si layers) and may require additional technological steps (e.g., alkaline mask removal, anodic bonding of Si plates for borosilicate glass etching). Cr/Au-based metal masks are the most commonly used in wet glass etching [12,18-22]. Chromium ensures high adhesion of gold films to glass, while gold is highly inert in HF solutions, which ensures deep microstructures etching. High cost and high diffusing ability of gold masks limit its possible applications. Refractory metals such as molybdenum and chromium are successfully used for deep glass etching [15-17, 24-26]. However, these metals tend to form high-stress layers requiring an advanced deposition process. Key benefits of molybdenum film are low dissolution rates in HF acid (near 19 Å/min) and high adhesion to the glass substrate [30], as well as lower cost compared to gold-based masks.

Physical properties of protective metal masks, substrate-metal interface and BOE solution concentration directly affect the quality of glass etching (fig. 1). Microdefects and metal films discontinuity cause pinhole defects on the glass surface (Fig. 1a). Poor adhesion results in protective mask undercutting and high lateral etch rates, (Fig 1b). High stress in the mask layers leads to the formation of microcracks (Fig. 1c) or increased roughness of etched edges (Fig. 1d). A high content of reaction etching products can lead to a substantial increase in the etched surface roughness (Fig. 1e) or even an irregular etching profile (Fig. 1f). High quality fused glass etching is characterized by both high etching rate and perfect surface quality, as well as a high etching isotropy. Etching solution optimization is one of the effective ways to improve etching quality, prevent defect formation and maintain mask stability.

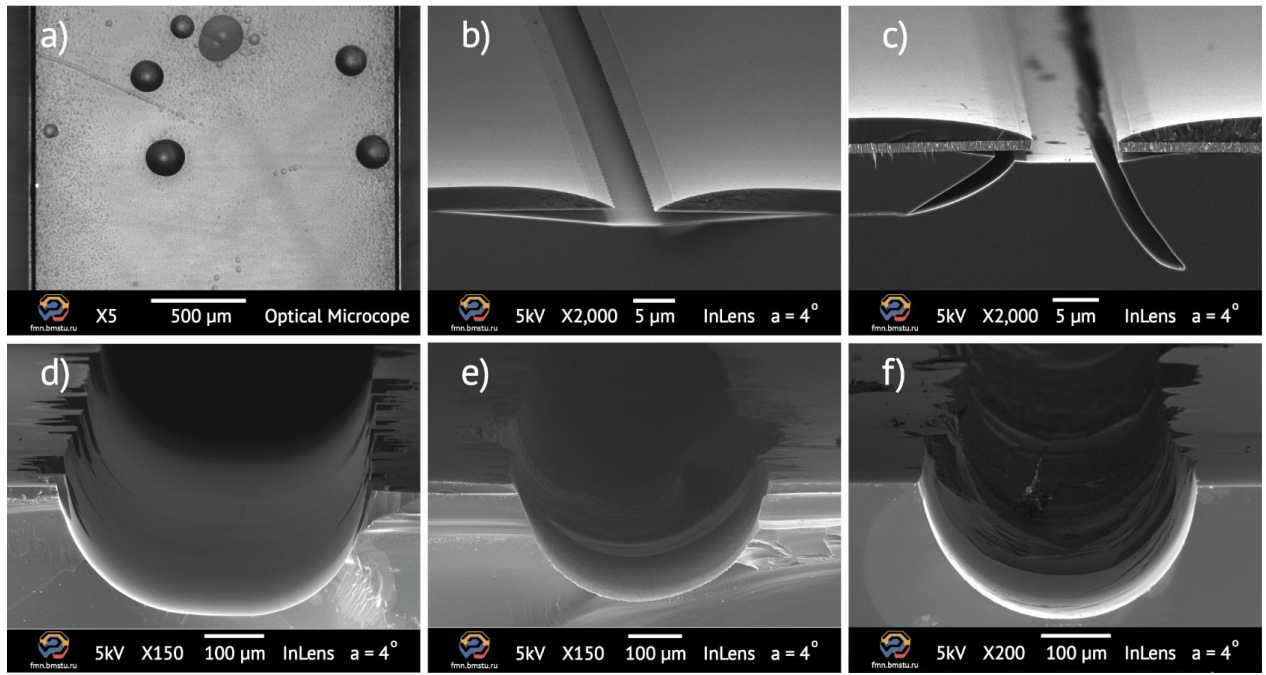

Fig. 1. Glass defects obtained after BOE etching process: a) optical image of pinholes on a glass surface and SEM images of: b) mask undercutting defect; c) microcracks of a glass substrate; d) rough structure edges; e) high surface roughness; f) irregular etching profile

In this work, we report on multilevel fused silica glass microstructures fabrication route based on over-200 μm deep wet etching through a stepped protective photoresist/metal mask with just a one grayscale photolithography step. There is trustworthy information on the effect of the concentration of the BOE solution on etched microstructures quality and mask resistance for deep glass etching process. For example, in [43], the influence of BOE solution concentration on $SiO_2$ etching profile shape for just a few micrometers depth is described. Our study reveals the dependence of the etching rate, depth, isotropy and mask stability on the composition of the BOE solution. Variation of BOE concentration leads to changing pH and amount of reaction species which affect the etching quality. The proposed etching process ensures mask stability, high etching rate and isotropy allowing to get multilevel etch profile in fused glass with a one initial photolithography step.

**2. Mechanism of a fused silica glass etching in hydrofluoric acid solutions**

Fused quartz is a pure silica glass ($SiO2$) in amorphous form. Breaking Si-O bonds is a critical aspect of the etching process since Si-O bonds have high bond strength (810 kJ/mol versus 327 kJ/mol Si-Si bond strength) [31]. Generally, the dissolution mechanism of $SiO2$ is described by three iterative stages: surface protonation, nucleophilic attack of the electrophilic silicon atom, and formation of the Si-F bond (Fig. 2). Depending on pH, there are two main effects in a solution: so-called surface and concentration effects [31, 32]. First, chemical equilibrium is established between the fluorine-containing components of the solution. Second, $SiOH^{2+}$, SiOH, or $SiO^-$ groups are formed on a glass surface due to a protonation or loss of a proton. The relative concentration of each of the three groups determines the reactivity of surface layers and depends on the pH of the solution. The number of SiOH groups prevails in the region of the $SiO_2$ isoelectric point pI (pI value for $SiO2$ varies from 2 to 4 [33]). Below the isoelectric point, the concentration of $SiOH^{2+}$ groups increases. Above it, the concentration of $SiO^-$ groups increases. The replacement of the $OH^–$ group from neutral SiOH is slower than the removal of $H2O$ from the protonated $SiOH^{2+}$ group [31].

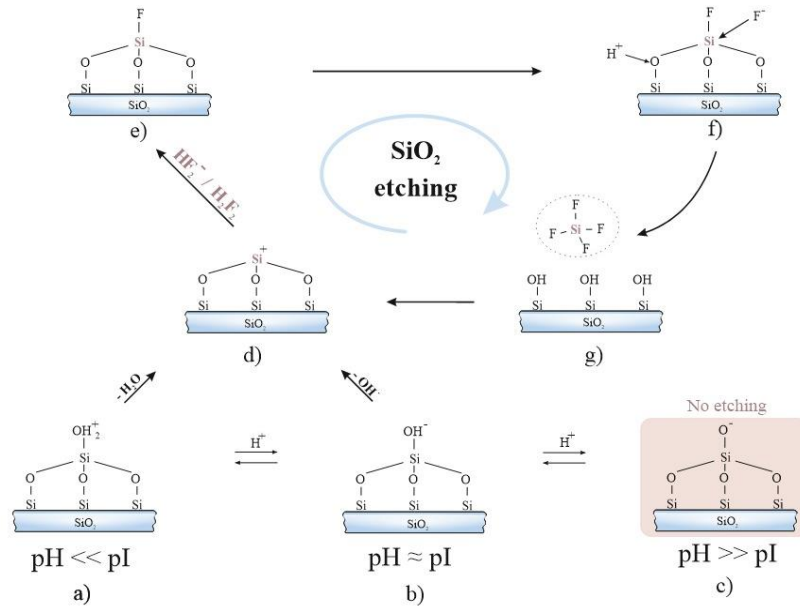

Fig. 2. Dissolution reaction mechanism of SiO₂ in HF-based solutions: a-c) equilibrium reaction of surface silanol groups in etched solution with different pH-value; d) nucleophilic attack of the electrophilic silicon atom; e-g) formation of the Si-F bonds.

The etching process is also influenced by etching solution composition, concentration of components, temperature, and mixing intensity of the solution. High process temperature increases the etching rate and decreases the probability of BOE solution crystallization during etching [34], but could worsen the etching isotropy [35]. Solution mixing only slightly affects the dissolution rate, because the process is kinetically controlled [32], but it can cause severe mask damage [25]. The etching rate increases with increasing HF content in the BOE solution. Adding an ammonium fluoride buffer NH4F to HF raises the etching resistance of photoresist masks and helps to maintain the etching rate [8], but its dependence on the NH4F content is non-linear. The etching rate increases with a small addition of NH4F to a certain concentration, but with a further NH4F concentration increase, it starts decreasing [36]. Thus, the etch rate depends on the percentage of etching particles in the solution, which can be described as solution pH. The pH of NH4F/HF systems is calculated according to (1) the Henderson-Hasselbalch equation [37].

$$pH = pK_a + lg\frac{[A^-]}{[HA]} \quad (1)$$

where $K_a$ is the dissociation constant of the weak acid, $pK_a = \log K_a$, and [HA] and [A⁻] are the molarities of the weak acid and its conjugate base.

There are many compounds described by (2)-(5) in BOE solution: HF, $F^-$, $(HF)_2$, $HF_2^-$, $NH_4^+$, $H^+$. In high HF-concentrated solutions the HF-based complexes like (HF)ₙF⁻ are formed [38]. $HF_2^-$ and $(HF)_2$ are the reaction species in HF solutions and the etching reaction rate with $HF_2^-$ is 2000-3000 times faster than with $(HF)_2$ [31,39], which can be explained by the bond angle of the species (180° and 90°, respectively) [31]. It has been shown that $F^-$ does not take part in etching reaction or, at least, that etching by $F^-$ is negligible [38]. The main reactions and equilibrium constants of reactions occurring in BOE solution, according to [40, 41]:

Equilibrium constant HF, k₁ = 6,9 · 10⁻⁴ mol/liter

$$HF \leftrightarrow H^+ + F^- \quad (2)$$

Equilibrium constant NH4F, k₂ = 44,17 mol/liter

$$NH_4F \leftrightarrow NH_4^+ + F^- \tag{3}$$

Complex formation HF, $kd_1 = 4{,}0$ mol$^{-1}$

$$HF + F^- \leftrightarrow HF_2^- \tag{4}$$

HF dimer formation, $kd_2 = 2{,}7$ mol$^{-1}$

$$HF + HF \leftrightarrow (HF)_2 \tag{5}$$

There several works describing calculations of reactive species percentage in BOE solution for different HF concentrations [31, 38, 39, 42], but they do not reflect the required range of BOE concentrations considered in this work. To describe the etching solution, the abbreviation "BOE n:m" will be used. This abbreviation refers to a mixing ratio of n parts by volume of 40% weight of NH4F and m parts 49% weight of HF. We calculated the relative content of components in BOE solution with the ratio n:m and solution pH using the reaction constants (2)-(5) (Fig. 3). These dependences are qualitative, since it do not take into account the formation of more complex particles and is calculated at room temperature, but it makes possible to visually understand the etching process.

Glass etching proceeds in the general case according to (6):

$$SiO_2 + 6HF \rightarrow H_2SiF_6 + 2H_2O \tag{6}$$

The reaction is multistage, proceeding according to (7) and (8):

$$SiO_2 + 4HF \rightarrow SiF_4 + 2H_2O \tag{7}$$

$$SiF_4 + 2HF \rightarrow H_2SiF_6 \tag{8}$$

Taking into account that $HF_2^-$ and $(HF)_2$ are reaction species in BOE solutions, (7) can proceed along two pathways (9) and (10). The dominant reaction depends on the particles percentage of each type in the solution, which in turn depends on the concentration of the initial reagents in the solution.

$$2H^+ + 2HF_2^- + SiO_2 \leftrightarrow SiF_4 + 2H_2O \tag{9}$$

$$2H_2F_2 + SiO_2 \rightarrow SiF_4 + 2H_2O \tag{10}$$

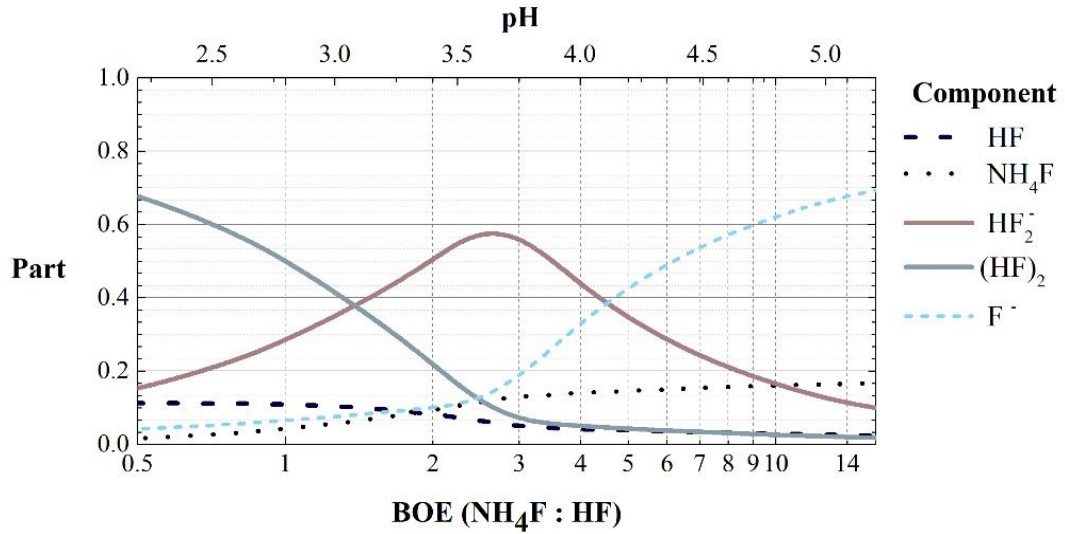

Fig. 3. The calculated fractions of main components in BOE solutions as a function of pH and BOE n:m ratio.

Thus, the fused silica etching rate will be dominantly depend on the concentration of the initial reagents in BOE solution. One can assume that the maximum etching rate will take place in the region close to the maximum concentration of $HF_2^-$.

## 3. Materials and methods

Fused silica glass 25 mm x 25 mm substrates of 500 μm thick are used in this study. The wafers were cleaned in organic solvents and sulfur-peroxide solutions (Piranha solution). A protective mask consisting of 200 nm thick molybdenum is sputtered on the top of glass wafer by magnetron sputtering at a base pressure of 3 mTorr. A 3 μm thick positive phorotesist (SPR220) was spin-coated on the wafers and patterned using standard photolithography process. Exposed areas of Mo were etched away in nitric, acetic and orthophosphoric acids solution. Subsequent heat treatment leads to strong crosslinking of the photoresist layer, which prevents the etching solution from penetrating deep into the mask.

Glass etching solution is freshly prepared by mixing hydrofluoric acid (HF 49%) and ammonium fluoride (NH4F 40%). The volume ratio (NH4F to HF) varies from 1 to 14 for glass etching. Buffered oxide etchant prevents strong penetration of the solution into the mask-substrate interface, pinhole formation and stabilizes etch rate. The process is carried out in a fluoroplastic tank with temperature control at a temperature of 60°C to increase the etching rate and prevent the crystallization of the solution. Test line structures with a width from 5 to 200 μm are further used to evaluate the etching process. The etching process was controlled by optical and scanning electron microscopy to assess the depth of etching and the quality of the surface after etching.

## 4. Results and discussion

After each etching process the photoresist and metal masks were removed, the samples were cleaved to evaluate the etching profile of the test structures. There is no data available for BOE 1:1 concentration, as the mask completely lost adhesion during etching. For BOE concentrations higher than 2.5, the mask withstood the etching process for 60 minutes. Fig. 4 demonstrates scanning electron microscope (SEM) images of etched test lines with a width of 50 μm for a different BOE concentration and 60 minutes etching time.

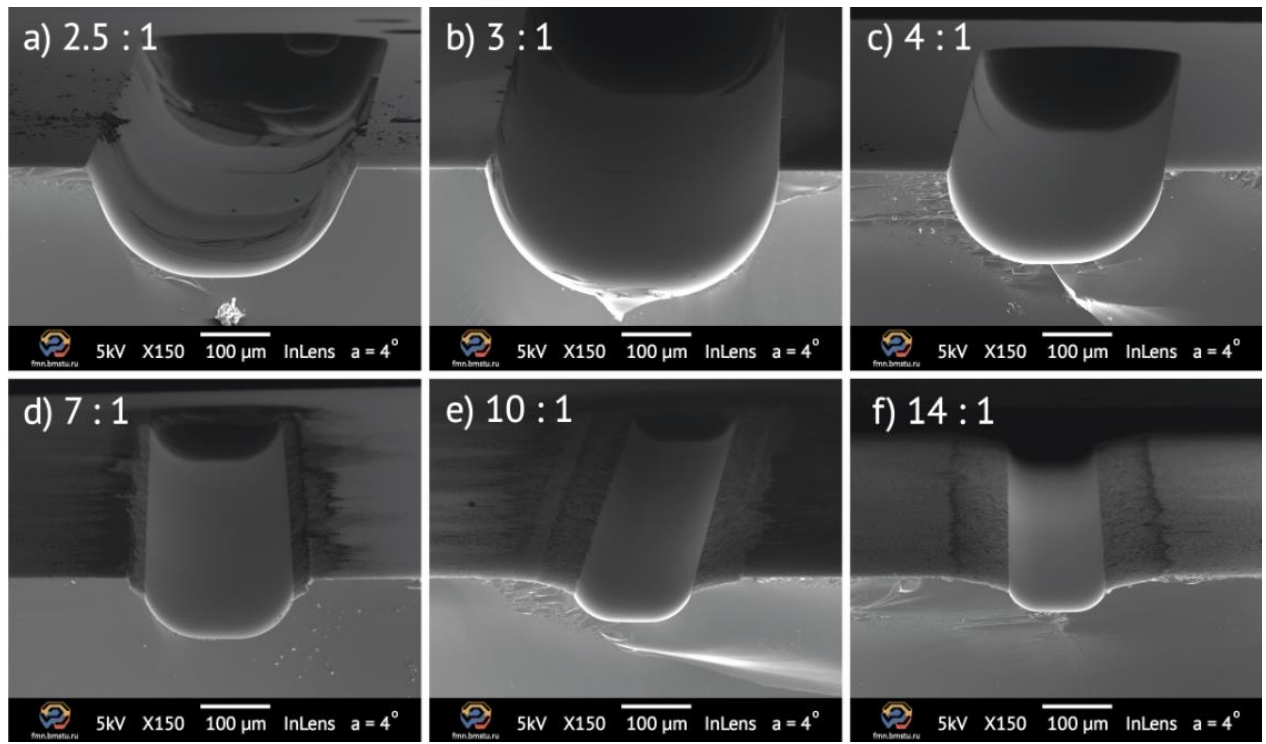

Fig. 4. SEM images of cleaved 50 μm wide test structures after wet etching and masks removal:
a) BOE 2.5:1 - non-isotropic etch region; b) BOE 3:1 and c) BOE 4:1 - isotropic etch region; d) BOE 7:1, e) BOE 10:1, and f) BOE 14:1 - mask undercutting etch region.

There are three typical regions that can be distinguished in term of etch profile shape: isotropic etching, non-isotropic etching and mask-undercutting region. Etching in solutions less than BOE 3:1 (fig. 4a) tends to form uneven etching bottom and non-isotropic profile characterized by higher lateral velocity then vertical one. For BOE 7:1 and higher (fig. 4c-e), we observed undercut defects through the mask with high roughness. Between these regions (fig. 4b, 4c) there is a defect-free etching, which is characterized by the absence of mask-undercutting and isotropic profile. BOE 3:1 and 4:1 provides smooth and defect-free surface. The maximum etch depth over-200 μm is obtained with BOE 3:1.

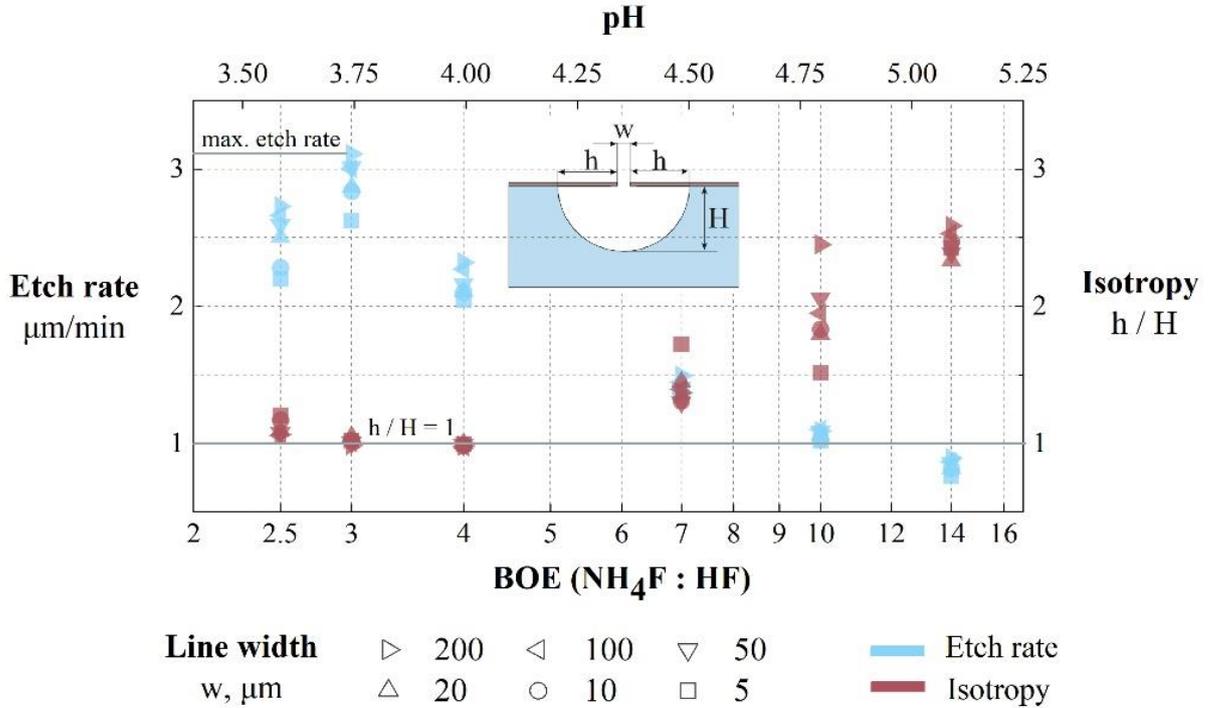

Fig. 5. The dependence on BOE concentration and pH value for different etched test line widths of fused silica etched profile isotropy (red dots), etch rate (blue dots), and preferred isotropy (solid line)

We demonstrate the dependencies (Fig. 5) of etch isotropy (the ratio of lateral undercut «h» to etching depth «H») and etch rate of fused silica on a BOE solution concentration (red and blue dots respectively). One can see that the maximum etch rate corresponds to BOE 3:1 with the maximum $HF_2^-$ concentration. The etch isotropy deviates from unity and the etch rate decreases for all the other concentrations, which occur due to chemical reactions in etch solution with its composition changing. For example, at high $NH_4F$ (BOE > 4:1) concentration dominating $F^-$ and $NH_4^+$ ions from the dissociation reaction of the $NH_4F$ buffer additive (2) negatively influences the etching process. $NH_4^+$ ions inactivate $HF_2^-$ leading to complex formation of a sparingly soluble $NH_4HF_2$ crystal. Besides, $NH_4^+$ ions passivate a negatively charged $SiO_2$ surface, preventing etching reaction by blocking deep penetration into $SiO_2$. Another negative effect is a precipitation of reaction products (11-12) due to its limited solubility [34].

$$SiO_2 + 3HF_3^- + H^+ \rightarrow SiF_6^{2-} + 2H_2O \qquad (11)$$

$$2NH_4^+ + SiF_6^{2-} \rightarrow (NH_4)_2SiF_6 \downarrow \qquad (12)$$

One can notice that the big amount of NH4F (BOE > 4:1) in solution decreases etching rate due to glass surface passivation with deposition of insoluble reaction products. In contrast, area with a high HF content (BOE < 3:1) are characterized by decrease in mask resistance and increase in the lateral etching rate. In common, a decrease of the etching rate is observed with a decrease in the width of the etched test lines. The etching rate for narrow test lines (5 μm) is 1.05–1.2 times lower than for the wider test lines (200 μm). It can be explained by the fact that in wide trenches all the reagents and reaction products are removed faster than the solution starts to deplete.

## 5. Multilevel fused silica glass microstructures fabrication route

We propose a fabrication route for multilevel glass structures including through-wafer holes and membranes in fused silica glass through a stepped mask. We used a 500-um thick fused silica UV-grade wafers (Siegert wafer), which were cleaned in a Piranha solution (H2SO4:H2O2) at 120°C. A 200 nm-thick molybdenum protective mask layer was magnetron sputtered at 3 mTorr process pressure on both sides of the wafers. Next, we relieved residual stresses in metal by post-thermal annealing with a temperature above 600°C in argon atmosphere. A stepped resist mask was patterned on glass wafers with a grayscale lithography (fig. 6a) in a spin-coated 5-um thick MEGAPOSIT SPR-220-7.0 photoresist layer (Micro resist technology GmbH, Germany). In the exposed area, a protective molybdenum film was wet etched in a mix of nitric, acetic and orthophosphoric acids (Transene company Inc.) at room temperature. Hard baking at 120°C was performed before wet etching the wafers to improve thermo-crosslinking-enhanced mechanical properties of the photoresist layer. Glass etching was carried out at 60°C in the proposed BOE 3:1 solution freshly prepared from NH4F 40% (Sigma Aldrich Inc.) and HF 49% (Technic Inc.) to form the firs etched level (fig. 6b). Further, we local stripped two-layer (resist-Mo) protective mask by oxygen plasma etching of photoresist thinner layer (fig. 6c) and followed molybdenum wet etching through open resist areas. The second glass etching step using the same procedure ensure forming the required microstructural multilevel mold. By varying the amount of grayscale steps, the number of etching steps and levels correspondingly can be increased.

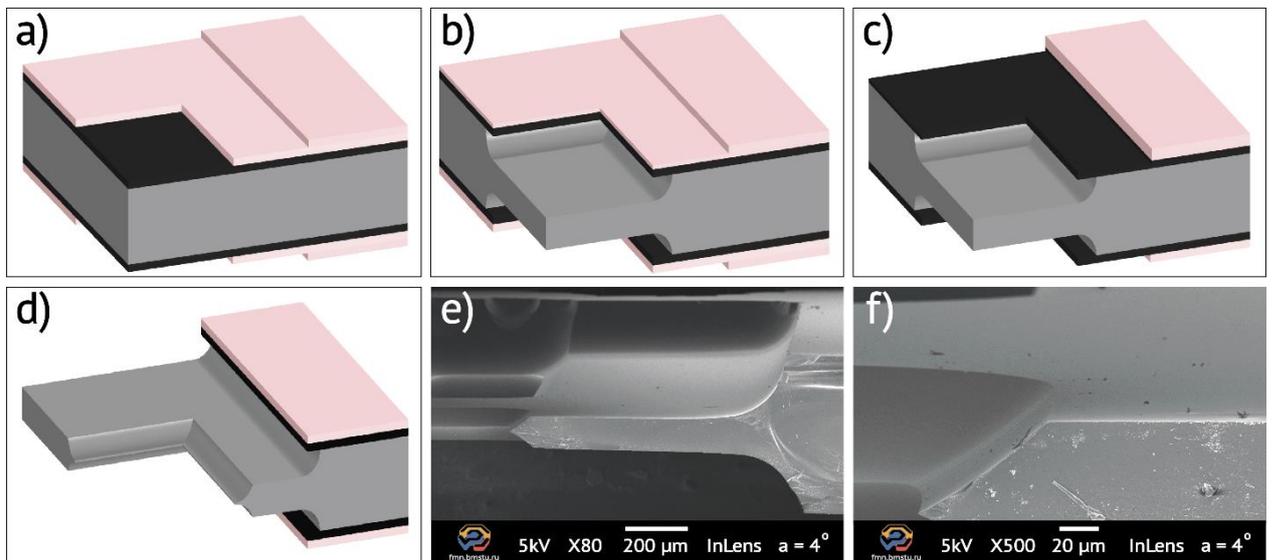

Fig. 6. Multilevel fused silica glass microstructures fabrication route : a) stepped profile resist patterning with grayscale photolithography; b) opening of the window in the metal mask and the first stage of etching; c) opening of the second stage mask; d) second pickling stage; e) SEM image of the cleaved membrane microstructure (etching result, x80k); f) SEM image of the cleaved membrane surface (x 500k).

The proposed method for processing fused silica glass and multilevel microstructures patterning clearly demonstrates the possibility of forming membranes, microchannels, and through-wafer holes with ultra-low roughness (fig. 6e, 6f). It opens the way to use the key advantages of fused silica glass, especially its thermal expansion coefficient, optical transparency, and high chemical inertness for fabrication microdevices with complex multilevel elements and multi wafer assembly.

## 6. Conclusion

In this work, we proposed a defect-free multilevel fused silica glass microstructures fabrication route based on deep wet etching in BOE solution through a stepped mask with grayscale photolithography. Fused silica glass is widely used in high quality factor MEMS devices due to stable properties under critical conditions. However, high-quality glass processing is possible in extremely reactive and dangerous solutions of hydrofluoric acid, which contributes to the formation of various defects. We reported a theoretical description of the etching process in terms of the reactions and the products formed. Then, composition of main fluoride components like $HF_2^-$, $F^-$, $(HF)_2$ as a function of BOE solution concentration and pH value were calculated for fused silica etching rate estimation. Based on our calculation and experiments we demonstrated that BOE concentration (NH$_4$F:HF from 1:1 to 14:1) directly determines the metal/photoresist mask resistance, etch rate and profile isotropy. We used molybdenum thin mask as a protective layer due to low dissolution rates in HF-based solutions and high adhesion to glass substrates. We confirmed our analytical evaluation with the experimental results by demonstrating the isotropy etching with smooth surface and maximum etching rate provided at BOE concentration of 3:1 (pH 3.75). It corresponds to calculated maximum of $HF_2^-$ content. Finally, we demonstrate a fabrication route based on high resistant protective mask for multilevel microelements patterning. It includes three key steps: one grayscale lithography step, wet etching in BOE 3:1 and ion-plasma photoresist thinning. The process ensures achieving a high-quality multilevel isotropic over-200 μm etching with the rate up to 3 μm/min for advanced fused silica microdevices with flexure suspensions, inertial masses, microchannels, and through-wafer holes. It should be noted that the proposed route assumes just a one initial photolithography step, which is critically important for advanced microdevices as it requires no mask formation on a non-planar prepatterned surfaces.

**Acknowledgements**

Technology was developed and samples were fabricated at the BMSTU Nanofabrication Facility (Functional Micro/Nanosystems, FMNS REC, ID 74300).